\documentclass[aps,prb,twocolumn,showpacs,groupedaddress,floatfix]{revtex4}

\usepackage{graphicx}
\usepackage{latexsym}
\usepackage{amssymb}

\begin{document}


\title{Measurement of mutual inductance from frequency dependence of impedance of AC coupled circuits using a digital dual-phase lock-in amplifier}


\author{Michael J. Schauber}
\affiliation{Department of Physics, State University of New York at Binghamton, Binghamton, New York 13902-6000}

\author{Seth A. Newman}
\affiliation{Department of Physics, State University of New York at Binghamton, Binghamton, New York 13902-6000}

\author{Lindsey R. Goodman}
\affiliation{Department of Physics, State University of New York at Binghamton, Binghamton, New York 13902-6000}

\author{Itsuko S. Suzuki}
\email[]{itsuko@binghamton.edu}
\affiliation{Department of Physics, State University of New York at Binghamton, Binghamton, New York 13902-6000}

\author{Masatsugu Suzuki}
\email[]{suzuki@binghamton.edu}
\affiliation{Department of Physics, State University of New York at Binghamton, Binghamton, New York 13902-6000}


\date{\today}

\begin{abstract}
We present a simple method to determine the mutual inductance $M$ between two coils in a coupled AC circuit by using a digital dual-phase lock-in amplifier. The frequency dependence of the real and imaginary parts is measured as the coupling constant is changed. The mutual inductance $M$ decreases as the distance $d$ between the centers of coils is increased. We show that the coupling constant is proportional to $d^{-n}$ with an exponent $n$ ($\approx$ 3). This coupling is similar to that of two magnetic moments coupled through a dipole-dipole interaction.
\end{abstract}

\pacs{01.50.Pa, 01.50.Qb, 01.40.Fk}

\maketitle



\section{\label{intro}Introduction}
Faraday's law of magnetic induction states that a changing magnetic flux through a coil of wire with respect to time will induce an EMF in the wire.  In a coupled circuit, there are two coils of wire; a primary coil that is connected in series with the voltage source, and a secondary coil that is not connected to any voltage source. The secondary coil receives energy only by induction. The EMF in the secondary coil affects the voltage across the primary coil due to reflected impedance.\cite{r01,r02}  

Here we present a simple method to measure the frequency dependence of the real and imaginary parts of the output voltage across a series resistance in the primary circuit of the coupled coils. Changing the separation distance between the two coils facing each other leads to the change in the mutual inductance. The present method allows one to obtain a great deal of data in a reasonably short period of time. A background for the AC analysis of the AC coupled circuit in the frequency domain is presented in Sec.~\ref{back}. The frequency dependence of the real and imaginary parts of output voltage is formulated and is simulated using Mathematica. Our experimental results are reported in Sec.~\ref{result}. The mutual inductance is determined as a function of the distance between the centers of two coils from the frequency dependence of the real and imaginary parts. We show that the mutual inductance changes with the distance $d$ as $d^{-n}$ where $n \approx 3$. This coupling is the same as that of magnetic moments which are coupled by a dipole-dipole interaction.

The detection technique of the signal in the present method is similar to that used in the nuclear magnetic resonance,\cite{r03,Ref01} where the behavior of the real part (the dispersion) and imaginary part (absorption) of the Bloch magnetic susceptibility can be measured as a function of frequency near the resonance frequency. 

\section{\label{back}BACKGROUND}
\subsection{\label{backA}Self inductance and mutual inductance of coils used in the present work}

\begin{figure}
\includegraphics[width=6.5cm]{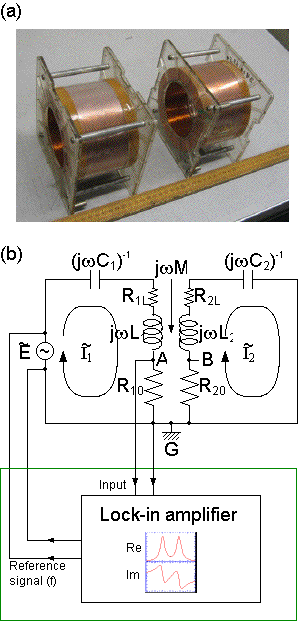}
\caption{\label{fig01}(Color online) (a) A picture of two coils used in the measurement. Each coil (Heath Company Part No. 40-694) has a cylindrical form with the inner diameter (9.30 cm), the outer diameter (12.5 cm), and the length (9.0 cm). The number of turns of the coil is $N=3400$. The value of $L_{1}=L_{2}$ are determined from the resonance frequency with $C_{1}=C_{2}=0.0038$ $\mu$F. The resonance frequency $f_{0}$ is equal to $f_{0}=1/(2\pi \sqrt{LC})=2850$ Hz. An AC voltage was supplied by the lock-in amplifier. (b) Frequency domain diagram of AC coupled circuit used in the present experiment. The primary circuit is on the left side, the secondary circuit on the right side. The coils 1 and 2 are used in place of the inductors $L_{1}$ and $L_{2}$. $L_{1}=L_{2}=0.821$ H. $C_{1}=C_{2}=0.0038$ $\mu$F. $R_{10}=R_{20}=31$ $\Omega$. $R_{1L}$ and $R_{2L}$ are resistances of coils 1 and 2. $R_{1L}=R_{2L}=66$ $\Omega$. $R_{1}=R_{10}+R_{1L}=97$ $\Omega$. $R_{2}=R_{20}+R_{2L}=97$ $\Omega$.}
\end{figure}

Figure \ref{fig01}(a) shows a photograph of coils which are used in the present work. For simplicity, we assume that each coil has a cylindrical form with a radius $R_{av}$ and a length $l$. The separation between the centers of two coils (see Fig.~\ref{fig01}(a)) is the distance $d$. The number of turns of the coils 1 and 2 is the same ($N$). The magnetic field produced by coil 1 (current $I_{1}$) at the center of coil 2 is given by
\begin{equation}
B=\frac{\mu_{0}}{2\pi} \frac{NA}{d^{3}}I_{1},
    \label{eq02}
\end{equation}
for $d \gg R_{av}$ using the Bio-Savart law.\cite{r01,r02} The voltage induced in the coil 2 is
\begin{equation}
V_{2}=-N\frac{d\Phi}{dt}=-NA\frac{dB}{dt}=-\frac{\mu_{0}N^{2}A^{2}}{2\pi d^{3}}\frac{dI_{1}}{dt},
    \label{eq03}
\end{equation}
where the magnetic flux is $\Phi =BA$ and $A$ is the cross-sectional area of the solenoid: $A =\pi R_{av}^{2}$. Thus the mutual inductance $M$ defined by $V_{2}= -MdI_{1}/dt$ is given by
\begin{equation}
M=\frac{\mu_{0}N^{2}A^{2}}{2\pi d^{3}}.
    \label{eq04}
\end{equation}
The ideal self-inductance $L_{0}$ is given by
\begin{equation}
L_{0}=\frac{\mu_{0}N^{2}A}{l},
    \label{eq05}
\end{equation}
for sufficiently long $l$, where $l$ is the length of solenoid, $\mu_{0}$ ($=4\pi \times 10^{-7}$ Tm/A) is a permeability, and $N$ is the total number of turn. Thus we have
\begin{equation}
k=\frac{Al}{2\pi d^{3}}=\frac{R_{av}^{2}l}{2d^{3}},
    \label{eq06}
\end{equation}
since $M=kL_{0}$ (see Eq.(\ref{eq12}) for general definition of $M$). The constant $k$ is dependent only on the geometry of the coils. Note that our coils used in the present measurement have $N$ = 3400 turn and $l$ =9.0 cm. When $R_{av}$ = 5.45 cm, we have $R_{av}^{2}l/2 = 133.7$, where $d$ is in the units of cm (see Fig.\ref{fig01}(a)). The self-inductance $L_{0}$ can be calculated as
\[
L_{0}=\frac{\mu_{0}N^{2}\pi R^{2}_{av}}{l}=1.506 \text{ H} .
\]
This value of $L_{0}$ is larger than the actual value of $L_{exp}$ ($= 0.821$ H, which will determined experimentally). The difference between $L_{exp}$ and $L_{0}$ is due to the deviation of the system from ideal one because of the finite length of the coil: $L_{exp} =K_{N}L_{0}$. $K_{N}$ is called the Nagaoka coefficient and is defined as\cite{Ref03,r06} 
\begin{equation}
K_{N}=\frac{4}{3\pi \sqrt{1-\alpha^{2}}}[\frac{1-\alpha^{2}}{\alpha^{2}}K(\alpha)+\frac{2\alpha^{2}-1}{\alpha^{2}}E(\alpha)-\alpha],
\label{eq07new}
\end{equation}
where
\begin{equation}
\alpha=\frac{1}{\sqrt{1+(l/2R)^{2}}},
\label{eq08new}
\end{equation}
$R$ and $l$ are the radius and length of coil, respectively, and $K(\alpha)$ and $E(\alpha)$ are the complete elliptic integral of the first and second kind,
\[
K(\alpha)=\int_{0}^{\pi /2}(1-\alpha^{2}\sin^{2}\theta)^{-1/2}d\theta ,
\]
and
\[
E(\alpha)=\int_{0}^{\pi /2}(1-\alpha^{2}\sin^{2}\theta)^{1/2}d\theta .
\]
The value of $K_{N}$ can be calculated as a function of $2R/l$. The ratio $K_{N}$ experimentally obtained in the present work is $L_{exp}/L_{0}$ = 0.821 H/1.506 H = 0.5458. This ratio is close to the ratio $K_{N}$ (= 0.6456) calculated from Eq.(\ref{eq07new}) for the ratio $2R_{av}/l=2\times 5.45/9.0\approx 1.211$.

\subsection{\label{backB}AC coupled circuit}
In Fig.~\ref{fig01}(b) we show the AC coupled circuit in the frequency domain.\cite{Ref04} The currents and voltages are all complex numbers. Using Kirchhoff's law, we can write down two equations,
\begin{equation}
\tilde{E}=\tilde{I}_{1}Z_{1}+j\omega M\tilde{I}_{2} ,
    \label{eq07}
\end{equation}
and
\begin{equation}
0=\tilde{I}_{2}Z_{2}+j\omega M\tilde{I}_{1} ,
    \label{eq08}
\end{equation}
where $j=\sqrt{-1}$, $\omega$ ($=2\pi f$) is the angular frequency, $\tilde{I}_{1}$ and $\tilde{I}_{2}$ are the loop currents of the primary and secondary circuit, $\tilde{E}$ is the source voltage, $Z_{1}$ and $Z_{2}$ are the impedance of the primary and secondary circuits, respectively,
\begin{eqnarray}
Z_{1}=R_{1} +jX_{1}, X_{1}=\omega L_{1}-\frac{1}{\omega C_{1}},\nonumber \\
Z_{2}=R_{2} +jX_{2}, X_{2}=\omega L_{2}-\frac{1}{\omega C_{2}},\nonumber
\end{eqnarray}
and $M$ is dependent on the distance between $L_{1}$ and $L_{2}$ coils. From Eqs.(\ref{eq07}) and (\ref{eq08}) we have
\begin{equation}
\tilde{E}=Z_{1}^{\prime}\tilde{I}_{1}, 
Z_{1}^{\prime}=Z_{1}+\frac{\omega^{2}M^{2}}{Z_{2}}, \label{eq09}
\end{equation}
where $Z_{1}^{\prime}$ is the effective impedance of the primary circuit. The effective impedance is rewritten as
\[
Z_{1}^{\prime}=R_{1}^{\prime}+jX_{1}^{\prime}
=(R_{1}+\frac{\omega^{2}M^{2}R_{2}}{R_{2}^{2}+X_{2}^{2}}) 
+j(X_{1}-\frac{\omega^{2}M^{2}X_{2}}{R_{2}^{2}+X_{2}^{2}}).
\]

For simplicity, we assume the symmetric configuration such that $R_{1}=R_{2}=R$, $C_{1}=C_{2}=C$, and $L_{1}=L_{2}=L$. Then we have $X=X_{1}=X_{2}=\omega L-1/\omega C$. The effective impedance $Z_{1}^{\prime}$ can be written as
\[
Z_{1}^{\prime}=R(1+\frac{\omega^{2}M^{2}}{R^{2}+X^{2}})+jX(1-\frac{\omega^{2}M^{2}}{R^{2}+X^{2}}).
\]
The voltage across $R_{10}$ between AG in Fig.~\ref{fig01}(a), $\tilde{V}_{R}$, is
\begin{equation}
\tilde{V}_{R}=\tilde{I}_{1}R_{10}=\tilde{E}\frac{R_{10}}{R}\tilde{G},
\end{equation}
where
\begin{equation}
\tilde{G}=\frac{R}{Z_{1}^{\prime}}=\frac{R}{R(1+\frac{\omega^{2}M^{2}}{R^{2}+X^{2}})+jX(1-\frac{\omega^{2}M^{2}}{R^{2}+X^{2}})}.
    \label{eq10}
\end{equation}
We define the ratio $x=\omega /\omega _{0}$, where $\omega _{0} =1/\sqrt{LC} $ and $x$ is always positive. The quality factor of the circuit is given by
\begin{equation}
Q=\frac{\omega_{0}L}{R} =\frac{1}{R}\sqrt{\frac{L}{C}} .
    \label{eq11}
\end{equation}
The mutual inductance $M$ is related to the self inductance $L$ by
\begin{equation}
M=k\sqrt{L_{1}L_{2}}=kL,
    \label{eq12}
\end{equation}
where $k$ is a constant and is smaller than 1. By using these relation, $C$ and $L$ can be expressed by $C=1/(\omega_{0}QR)$ and $L=QR/\omega_{0}$. Then 
$\tilde{G}$ can be rewritten as
\begin{widetext}
\begin{eqnarray}
\tilde{G}=\mu+j\nu ,     \label{eq13} \\
\mu =\frac{x^{2}\lbrack Q^{2}(1+k^{2})x^{4}+(1-2Q^{2})x^{2}+Q^{2}\rbrack}{x^{4}
+Q^{4} \lbrack (k^{2} -1)x^{4} +2x^{2} -1\rbrack^{2} +2Q^{2} \lbrack (1+k^{2}
)x^{6} -2x^{4} +x^{2}\rbrack} ,
    \label{eq14} \\
\nu =\frac{-Qx(x+1)(x-1)\lbrack Q^{2}(1-k^{2})x^{4} +(1-2Q^{2})x^{2} +Q^{2}\rbrack}{x^{4} +Q^{4} \lbrack (k^{2} -1)x^{4} +2x^{2} -1\rbrack^{2} +2Q^{2} \lbrack (1+k^{2} )x^{6} -2x^{4} +x^{2} \rbrack}, 
    \label{eq15}
\end{eqnarray}
\end{widetext}
which depends only on $x$, $Q$, and $k$. 
According to Eq.(\ref{eq15}), $\nu$ becomes zero when $x$ = 0 and 1. The other possible $x$'s for giving $\nu=0$ ca be examined by the following quadratic equation: 
\begin{equation}
Q^{2} (1-k^{2} )x^{4} +(1-2Q^{2} )x^{2} +Q^{2} =0 .
    \label{eq16}
\end{equation}
The solution of this equation is formally given by
\begin{equation}
x_{1}=\sqrt{\frac{2Q^{2}-1+\sqrt{1-4Q^{2}+4k^{2}Q^{4}}}{2(1-k^{2})Q^{2}}} ,
    \label{eq17}
\end{equation}
and
\begin{equation}
x_{2}=\sqrt{\frac{2Q^{2}-1-\sqrt{1-4Q^{2}+4k^{2} Q^{4}}}{2(1-k^{2})Q^{2}}}  ,
    \label{eq18}
\end{equation}
where
\begin{equation}
x_{1}x_{2}=\frac{1}{\sqrt{1-k^{2}}}  ,
    \label{eq19}
\end{equation}
and
\[
x_{1}^{2}-x_{2}^{2}=\frac{\sqrt{1-4Q^{2}+4k^{2}Q^{2}}}{Q^{2}(1-k^{2})}  .
\]
The values of $\mu$ and $\nu$ depend on $x$ and are described as coordinates in the $(\mu, \nu)$ plane for convenience; $(\mu, \nu)=(0,0)$ at $x=0$ and $\infty$, $(1/2,0)$ at $x=x_{1}$, and $(1/(1+k^{2}Q^{2}),0)$ at $x=1$. 

\begin{figure}
\includegraphics[width=7.0cm]{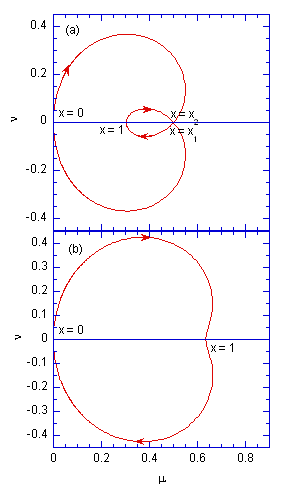}
\caption{\label{fig02}(Color online) (a) State-I. Simulation plot of the trajectory denoted by the point $(\mu,\nu)$ for $kQ >1$ and $L_{1}=L_{2}$ (the symmetric configuration), when $x$ ($= f/f_{0}$) varies from $x$ = 0 to $\infty$. The figure corresponds to the case of $kQ$ = 1.515 where $Q$ = 151.5 and $k$ = 0.01. The point on the trajectory is located at $(\mu,\nu)$ = (0, 0) for $x=0$, at (1/2, 0) for $x=x_{2}$, at ($1/[1+(kQ)^{2}]$, 0) for $x$ = 1, at (1/2, 0) for $x=x_{1}$, and at (0, 0) for $x = \infty$. (b) State-II. Simulation plot of the trajectory denoted by the point $(\mu,\nu)$ for $kQ<1$, when $x$ varies from $x$ = 0 to $\infty$. The figure corresponds to the case of $kQ$ = 0.7575 where $Q$ = 151.5 and $k$ = 0.005. The point is located at $(\mu,\nu)$ = (0, 0) for $x$ = 0, at ($1/[1+(kQ)^{2}]$, 0) for $x$ = 1, and at (0, 0) for $x = \infty$.}
\end{figure}

We now consider only the case of $k<1$ and $Q \gg 1$, which corresponds to the present experiment. 

(1) State-I: $4Q^{2}(k^{2}Q^{2}-1)+1>0$ . This condition is nearly equivalent to $kQ > 1$ since $Q \gg 1$. There are two solutions $x_{1}$ and $x_{2}$ (0$<x_{2}<1<x_{1}$) besides $x = 0$ and 1. In Fig.~\ref{fig02}(a) we show the simulation plot of the trajectory of the point ($\mu$,$\nu$) for $kQ > 1$ and $L_{1}=L_{2}$ (the symmetric configuration), when $x$ varies from $x = 0$ to $\infty$. This figure corresponds to the case of $kQ$ = 1.515 where $Q$ = 151. 5 and $k$ = 0.01.

(2) State-II: $4Q^{2} (k^{2} Q^{2} -1)+1<0$ . This condition is nearly equivalent to $kQ<$1. There is no solution, besides $x$ = 0 and 1. In Fig.~\ref{fig02}(b) we show the simulation plot of the trajectory of the point $(\mu,\nu)$ for $kQ < 1$ and $L_{1}=L_{2}$, when $x$ varies from $x$ = 0 to $\infty$. This figure corresponds to the case of $kQ$ = 0.758, where $Q$ = 151.5 and $k$ = 0.005.

\begin{figure}
\includegraphics[width=7.0cm]{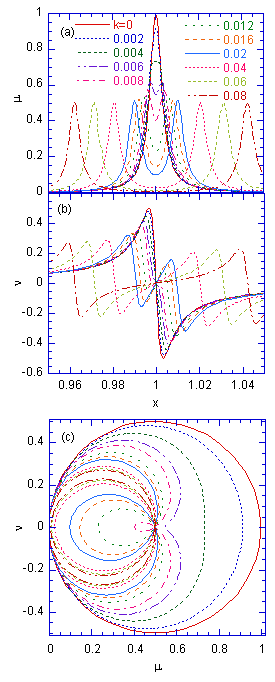}
\caption{\label{fig03}(Color online) Simulation plot of (a) the real part ($\mu$) and (b) the imaginary part ($\nu$) as a function of $x$ (calculations), where $Q$ = 151.5 and $L_{1}=L_{2}$ (symmetrical case). $k$ is changed as a parameter: $k$ = 0 - 0.08. (c) Typical trajectory denoted by the point $(\mu,\nu)$ for $Q$ = 151.5 and $L_{1}=L_{2}$, when $x$ is varied from $x$ = 0 to $\infty$. $k$ is changed as a parameter: $k$ = 0 - 0.08.}
\end{figure}

Figures \ref{fig03}(a) and (b) show simulated plots of the real part ($\mu$) and the imaginary part ($\nu$) as a function of $x$, where $Q$ = 151.5 and $L_{1}=L_{2}$. The coupling constant $k$ is changed as a parameter: $k$ = 0 - 0.08. In Fig.~\ref{fig03}(a), the double peaks in the $\mu$ vs $x$ curve are symmetric with respect to $x$ = 1 and become closer and closer together as $k$ is decreased and become a single peak at $kQ \approx 0.5757$. The imaginary part $\nu$ (see Fig.~\ref{fig03}(b)) has a positive local minimum at $x$ = 0.9967 and a negative local maximum at $x$ = 1.0033 in the limit of $k \rightarrow 0.$ Figure \ref{fig03}(c) shows the trajectory denoted by the point ($\mu$,$\nu$) for $Q$ = 151.5 and $L_{1}=L_{2}$ (the symmetric configuration), when $x$ is varied from $x$ = 0 to $\infty$. The coupling constant $k$ is changed as a parameter: $k$ = 0 - 0.1. There is a drastic change of the trajectory from the state-I and state-II at $kQ = 1$, when $kQ$ is decreased. For $k$ = 0, the trajectory is a circle of radius 1/2 centered at $\mu$ = 1/2 and $\nu$ = 0. 

\section{\label{exp}EXPERIMENTAL PROCEDURE}
The AC coupled circuit in the frequency domain was configured in Fig.~\ref{fig01}(b). An AC voltage was placed across the primary circuit, $v_{s}(t)=Re \lbrack \tilde{E}e^{j\omega t}\rbrack$, where $\tilde{E}$ is the complex voltage source. Two $LCR$ circuits are placed together with inductors, resistors and capacitors. In Fig.~\ref{fig01}(a) we show the overview of two coils used in the present measurement. The minimum distance between the centers of the coils is 10.2 cm. The two inductors are placed on a meter long track. 

We used a digital dual-phase lock-in amplifier (Stanford, SR850),\cite{Ref02} which is set at both a dual-phase mode and a frequency-scan mode. The real part $\mu$ (the in-phase signal), and imaginary part $\nu$ (the out-of-phase signal) were measured simultaneously, when the frequency was continuously changed across the desired frequency range. The frequency range initially chosen was 2500 Hz to 3500 Hz. At the 25.2 cm separation the frequency scan was only needed to be between 2700 Hz and 3000 Hz. The measurements are repeated for numerous distances ranging from $d=10.2$ to 70.2 cm. The output voltage across the resistance (AG) in Fig.~\ref{fig01}(b) is given by 
\begin{eqnarray}
V_{out}(t) &=& Re[\tilde{V}_{R}e^{j\omega t}] \nonumber\\
 &=& Re[(\mu + j\nu)E_{0}^{\prime}\exp [j(\omega t+\phi_{0})] \nonumber\\
 &=& E_{0}^{\prime}\mu \cos(\omega t+\phi_{0})+E_{0}^{\prime}\nu \cos(\omega t+\phi_{0}+\pi/2),\nonumber\\
    \label{eq01}
\end{eqnarray}
where $\tilde{E}=E_{0}e^{j\phi_{0}}$, $\phi_{0}$ is the phase, and $E_{0}^{\prime}=E_{0}R_{10}/R$. The in-phase component of the lock-in amplifier is equal to $E_{0}^{\prime}\mu$ and the out-of phase component is equal to $E_{0}^{\prime}\nu$, where $E_{0}^{\prime} = 4.97$ mV. Consequently, one can determine the values of the real part $\mu$ and the imaginary part $\nu$ defined by Eqs.(\ref{eq14}) and (\ref{eq15}), independently.   

\section{\label{result}RESULTS}

\begin{figure}
\includegraphics[width=7.0cm]{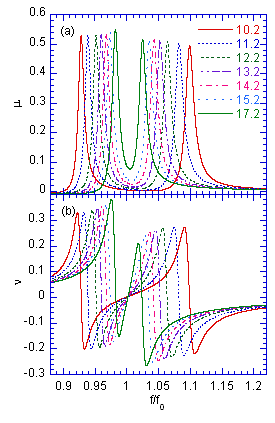}
\caption{\label{fig04}(Color online) Experimental plot of (a) the real part ($\mu$) and (b) the imaginary part ($\nu$) as a function of $x$ ($=f/f_{0}$) (in the frequency scan). The distance $d$ between the centers of two coils is changed as a parameter. $d$ = 10.2 cm - 17.2 cm. $f_{0}$ = 2850 Hz. $Q$ = 151.5. }
\end{figure}

\begin{figure}
\includegraphics[width=7.0cm]{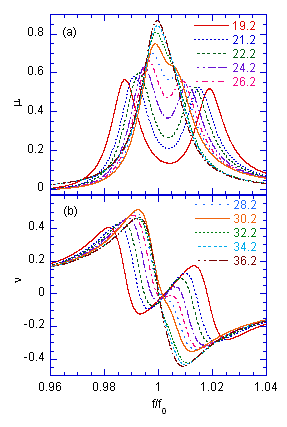}
\caption{\label{fig05}(Color online) Experimental plot of (a) the real part ($\mu$) and (b) the imaginary part ($\nu$) as a function of $x$ ($=f/f_{0}$) (in the frequency scan). The distance $d$ between the centers of two coils is changed as a parameter. $d$ = 19.2 cm - 36.2 cm. $f_{0}$ = 2850 Hz. $Q$ = 151.5.}
\end{figure}
 
\begin{figure}
\includegraphics[width=8.0cm]{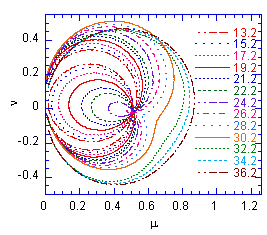}
\caption{\label{fig06}(Color online) Experimental trajectories of the point ($\mu$,$\nu$) for $Q$ = 151.5, when $x$ ($= f/f_{0}$) is varied from $x$ = 0.947 ($f$ = 2700 Hz) to 1.053 ($f$ = 3000 Hz). $f_{0}$ = 2850 Hz. The distance is changed as a parameter: $d$ = 13.2 - 36.2 cm. The deviation of the experimental trajectories from the ideal one as shown in Fig.~\ref{fig03}(c) is partly due to the asymmetric configuration ($L_{1}$ is slightly larger than $L_{2}$).}
\end{figure}
 
\begin{figure}
\includegraphics[width=7.0cm]{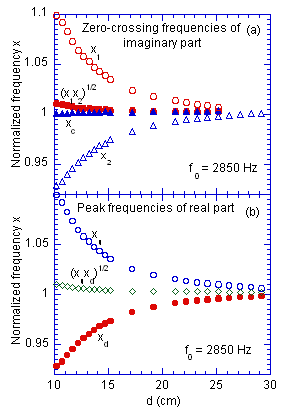}
\caption{\label{fig07}(Color online) The normalized zero-crossing frequencies of the imaginary part ($\nu$) (at which $\nu$ becomes zero) as a function of the distance $d$ (cm). $x_{2}=f_{2}/f_{0}$ ($<1$), $x_{c}=f_{c}/f_{0}$ ($\approx 1$), $x_{1}=f_{1}/f_{0}$ ($>1$), and $f_{0}$ = 2850 Hz. The normalized frequency defined by $(x_{1}x_{2})^{1/2}$ is also shown for comparison. (b) The normalized peak frequencies of the real part ($\mu$) as a function of the distance $d$ (cm). $x_{d}=f_{d}/f_{0}$ ($<1$) and $x_{u}=f_{u}/f_{0}$ ($>1$). $f_{0}$ = 2850 Hz. The real part $\mu$ takes two peaks at the lower and upper frequencies $f_{d}$ and $f_{u}$ for $d < 30$ cm. The normalized frequency defined by $(x_{u}x_{d})^{1/2}$ is also shown for comparison.}
\end{figure}

We have measured the frequency dependence of the real part $\mu$ and the imaginary part $\nu$ when the distance $d$ between the centers of two coils is changed as a parameter: $d$ = 10.2 - 70.2 cm. Our results are shown in Figs.~\ref{fig04} - \ref{fig06}. Figures \ref{fig04} and \ref{fig05} show the experimental plots of the real part ($\mu$) and the imaginary part ($\nu$) as a function of $x$ ($= f/f_{0}$), where $f_{0}$ = 2850 Hz and $Q$ = 151.5. In Figs.~\ref{fig04}(a) and (b) double peaks of $\mu$ become closer and closer when $d$ is increased and the double peaks become a single peak around $d$ = 36.2 cm. The double peaks are not symmetric with respect to $x$ = 1. The peak at the lower-$x$ side is higher than that at the higher-$x$ side. The real part has a local minimum at $x$ which is a little larger than 1. In Fig.~\ref{fig05}(a) and (b) the imaginary part $\nu$ crosses the $\nu = 0$ line at $x=x_{2}$, $x \approx 1$, and $x=x_{1}$ ($> x_{2}$). The positions $x_{1}$ and $x_{2}$ become closer and closer as $d$ is increased and combine into the position $x \approx 1$, but not at $x$ = 1. Figure \ref{fig06} shows the experimental trajectories of the point $(\mu,\nu)$, when $x$ ($= f/f_{0}$) is varied from $x$ = 0.947 ($f$ = 2700 Hz) to 1.053 ($f$ = 3000 Hz). The distance $d$ is changed: $d = 13.2 - 36.2$ cm. The transition occurs between the state-I and state-II at $d \approx 27.2$ cm. The overview of our trajectory is similar to the simulation plot as shown in Fig.~\ref{fig03}(c). However our trajectory rotates clockwise compared to the ideal simulation plot (the symmetric configuration). The deviation of our trajectories from the ideal case (Fig.~\ref{fig03}(c)) is partly due to the asymmetric configuration ($L_{1}$ is slightly larger than $L_{2}$).

Figure \ref{fig07}(a) shows the zero-crossing frequencies normalized by $f_{0}$ for the imaginary part ($\nu$) (at which $\nu$ becomes zero) as a function of the distance $d$ (cm), where $x_{2}=f_{2}/f_{0}$ ($<1$), $x_{c}=f_{c}/f_{0}$ ($\approx 1)$, and $x_{1}=f_{1}/f_{0}$ ($> 1$). The value of $x_{c}$ is a little different from 1. In Fig.~\ref{fig07}(a) we also show the normalized frequency defined by $(x_{1}x_{2})^{1/2}$. This frequency decreases with increasing $d$. This implies that the parameter $k$ decreases with increasing $d$ as predicted from Eq.(\ref{eq19}).
 
Figure \ref{fig07}(b) shows the normalized peak frequencies of the real part ($\mu$) as a function of the distance $d$ (cm), where $x_{d}=f_{d}/f_{0}$ ($<1$) and $x_{u}=f_{u}/f_{0}$ ($>1$). The real part $\mu$ takes double peaks at the lower and upper frequencies $f_{d}$ and $f_{u}$ for $d<30$ cm. The $d$ dependence of $x_{u}$ and $x_{d}$ is similar to that of $x_{1}$ and $x_{2}$, respectively. In Fig.~\ref{fig07}(b) we also show the normalized frequency defined by $(x_{u}x_{d})^{1/2}$ as a function of $d$. This frequency decreases with increasing $d$ like $(x_{1}x_{2})^{1/2}$ in Fig.~\ref{fig07}(a).

\section{\label{dis}DISCUSSION}
First we show that, from a view point of physics, the AC coupled circuit with the mutual inductance is equivalent to the magnetic moments $\tilde{M}_{1}$ ($= \mu_{0}NI_{1}A$) for the coil 1 and $\tilde{M}_{2}$ ($=\mu_{0}NI_{2}A$) for the coil 2. They are coupled with a dipole-dipole interaction defined by
\begin{equation}
U_{12}=\frac{1}{4\pi\mu_{0}}\lbrack \frac{{\bf \tilde{M}}_{1} \cdot {\bf \tilde{M}}_{2}}{r^{3}}-\frac{3({\bf \tilde{M}}_{1}\cdot {\bf r})({\bf \tilde{M}}_{2}\cdot {\bf r})}{r^{5}} \rbrack ,
    \label{eq20}
\end{equation}
where ${\bf r}$ is the position vector connecting the centers of coils 1 and 2. When both ${\bf \tilde{M}}_{1}$ and ${\bf \tilde{M}}_{2}$ are parallel to the direction of ${\bf r}$, a parallel alignment of two magnetic moments is energetically favorable,
\begin{equation}
\tilde{U}_{12}=-\frac{2}{4\pi \mu_{0}}\frac{\tilde{M}_{1}\tilde{M}_{2}}{d^{3}}
=-\frac{\mu_{0}^{2}N^{2}A^{2}}{2\pi \mu_{0}d^{3}}I_{1}I_{2}.
    \label{eq21}
\end{equation}
From the definition of the mutual inductance $M$, the interaction energy $\tilde{U}_{12}$ can be described by $\tilde{U}_{12}=-MI_{1}I_{2}$, leading to the mutual inductance which is the same as Eq.(\ref{eq04}) derived from Faraday's law.

\begin{figure}
\includegraphics[width=7.0cm]{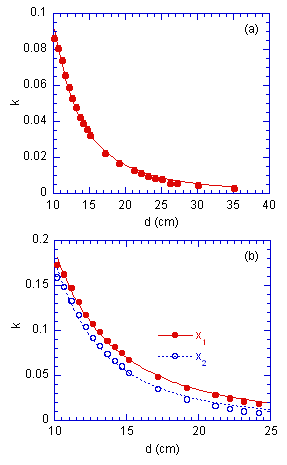}
\caption{\label{fig08}(Color online) (a) and (b) Plot of $k$ as a function of the distance $d$. $Q$ = 151.5. (a) The value of $k$ is derived from the prediction that the real part ($\mu$) is equal to $1/[1+(kQ)^{2}]$ at $x$ = 1 for the symmetrical configuration ($L_{1}=L_{2}$). The best fitted curve to the expression given by Eq.(\ref{eq22}) is denoted by a solid line. (b) The values of $k$ are derived from the prediction that $x_{1}$ and $x_{2}$ are described by Eqs.(\ref{eq17}) and (\ref{eq18}). The values of $k$ are numerically solved for each $d$. The best fitted curves are shown by the dotted and solid lines in the figure.}
\end{figure}

It is predicted that the parameter $k$ changes with distance according to Eq.(\ref{eq06}); $k$ is proportional to $d^{-3}$. The first method to determine the parameter $k$ as a function of the distance $d$, is as follows. As shown in Sec.~\ref{back}, it is predicted that in the symmetrical configuration ($L_{1}=L_{2}=L$), the real part $\mu$ takes a $\mu = 1/[1+(kQ)^{2}]$ at $x$ = 1. Note that the imaginary part $\nu$ is equal to zero at $x$ = 1. Experimentally we determine the value of $\mu$ at $x$ = 1 as a function of $d$. The values of $k$ are derived from the above expression with $Q$ = 151.5. Figure \ref{fig08}(a) show the plot of $k$ vs $d$ thus obtained. The value of $k$ drastically decreases with increasing $d$ and almost reduces to zero at $d$ = 35 cm. The least-squares fit of the data of $k$ vs $d$ to an expression 
\begin{equation}
k=\frac{\zeta}{d^{n} },
    \label{eq22}
\end{equation}
yields a constant $\zeta = 38.5 \pm 5.0$ and the exponent $n = 2.60 \pm 0.05$, where $d$ is in the units of cm. The value of $\zeta$ is rather different from the predicted value for the present coils ($\zeta$ = 151.5), while the value of $n$ is rather close to the predicted value ($n$ = 3). The large deviation of the experimental value of $\zeta$ from our prediction may be related to the asymmetric configuration of $L_{1}$ and $L_{2}$ in the present system, where $L_{2}$ is slightly lower than $L_{1}$ (which will be discussed later). As shown in Fig.~\ref{fig04}(a), the value of $x$ where the real part $\mu$ has a local minimum is not equal to $x$ = 1, and shifts to the high-$x$ side.

The second method to determine the value of $k$ is as follows. In Sec.~\ref{back}, it is predicted that the imaginary part $\nu$ takes zero-crossing at $x=x_{2}$, 1, and $x_{1}$ in the case of the symmetrical configuration ($L_{1}=L_{2}=L$). Note that the imaginary part $\nu$ is not always equal to zero at $x$ = 1 partly because of the asymmetric configuration in the present experiment. The value of $k$ for each $d$ is derived by applying the Mathematica program called "FindRoot" to Eq.(\ref{eq17}) with the experimental value of $x_{2}$ and to Eq.(\ref{eq18}) with the experimental value of $x_{1}$ (see Fig.~\ref{fig07}(a)), since Eqs.(\ref{eq17}) and (\ref{eq18}) are complicated functions of $k$. In Fig.~\ref{fig08}(a) we show the value of $k$ as a function of $d$ thus obtained. The value of $k$ drastically decreases with increasing $d$. The value of $k$ is a little larger than those obtained from the first method at the same $d$. The least-squares fit of the data of $k$ vs $d$ to Eq.(\ref{eq22}) yields the parameters $\zeta =57.1 \pm 6.9$ and $n=2.48 \pm 0.05$ for $x_{1}$ and $\zeta =149 \pm 33$ and $n=2.92 \pm 0.09$ for $x_{2}$. The latter result is in excellent agreement with the prediction ($\zeta$ = 151.5 and $n$ = 3.0). Such different values of $\zeta$ are partly due to the effect of the asymmetric configuration of coils. Nevertheless, it may be concluded experimentally that two magnetic moments made from coils are coupled through the dipole-dipole interaction with the exponent $n$ being equal to 3.

\begin{figure}
\includegraphics[width=7.0cm]{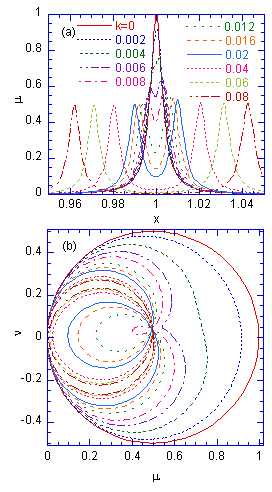}
\caption{\label{fig09}(Color online) (a) Simulation plot of the real part ($\mu$) as a function of $x$, where $Q$ = 151.5 and $L_{1} = 0.8207$ H and $L_{2} = 0.8215$ H (the asymmetric configuration). (b) The trajectory of the point ($\mu$,$\nu$) for $Q$ = 151.5, $L_{1} = 0.8207$ H and $L_{2} =0.8215$ H, when $x$ is varied from $x$ = 0 to $\infty$. The coupling constant $k$ is changes as a parameter: $k$ = 0 - 0.08.}
\end{figure}

\begin{figure}
\includegraphics[width=7.0cm]{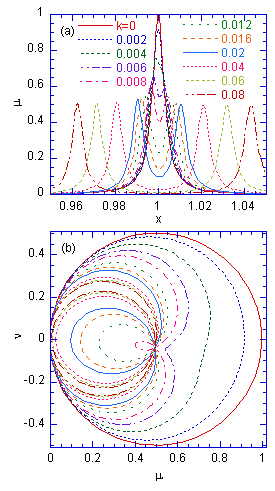}
\caption{\label{fig10}(Color online) (a) Simulation plot of the real part ($\mu$) as a function of $x$, where $Q$ = 151.5 and $L_{1} = 0.8207$ H and $L_{2} = 0.8198$ H. The coupling constant $k$ is changed as a parameter: $k$ = 0 - 0.08.(b) Typical trajectory denoted by the point ($\mu$,$\nu$) for $Q$ = 151.5, when $x$ is varied from $x$ = 0 to $\infty$. The coupling constant $k$ is changes as a parameter: $k$ = 0 - 0.08.}
\end{figure}

Finally we discuss the effect of the asymmetric configuration on the trajectory in the ($\mu$,$\nu$) plane. As shown in Fig.~\ref{fig06}, the trajectory rotates clockwise compared to the case of the trajectory in the symmetric configuration. Figure \ref{fig09}(a) shows the simulation plot of the $\mu$ as a function of $x$ for the asymmetric configuration ($L_{1}=0.8201$ H and $L_{2}=0.8215$ H) as $k$ is changed as a parameter. Double peaks of $\mu$ around $x$ = 1 are not symmetric with respect to $x$ = 1. The peak at the high-$x$ side is higher than that at the low-$x$ side. Double peaks become closer and closer as $k$ is decreased. Figure \ref{fig09}(b) shows the simulation plot of the trajectory in the ($\mu$,$\nu$) plane under the same condition as Fig.~\ref{fig09}(a). The trajectory rotates counterclockwise compared to the case of the trajectory in the symmetric configuration. Figures \ref{fig10}(a) and (b) show the simulation plot of $\mu$ as a function of $x$ and the trajectory in the ($\mu$,$\nu$) plane for the asymmetric configuration ($L_{1} = 0.8201$ H and $L_{2} = 0.8198$ H) where $k$ is changed as a parameter. Double peaks of $\mu$ around $x$ = 1 are not symmetric with respect to $x$ = 1. The peak at the high-$x$ side is lower than that at the low-$x$ side. The trajectory rotates clockwise compared to the case of the trajectory in the symmetric configuration. These features are in good agreement with those observed in the present measurement (see Fig.~\ref{fig05}(a) for the $\mu$ vs $x$ curve and Fig.~\ref{fig06} for the trajectory). So we can conclude that $L_{1}$ is a little larger than $L_{2}$, which means the asymmetric configuration for the present measurement.

\section{CONCLUSION}
We present a simple method for determining the mutual inductance of the AC coupled circuit using a digital dual-phase lock-in amplifier. This method allows one to get a large amount of data on the frequency dependence of the real and imaginary part of the AC output voltage in a reasonably short time. Our experimental results show that the coupling constant of the two coils is proportional to $d^{-n}$ with an exponent $n$ ($\approx 3$), where $d$ is the distance between the centers of coils. This coupling is similar to that of two magnetic moments coupled through a dipole-dipole interaction.

\begin{acknowledgments}
We are grateful to Mark Stephens for providing us with two coils with almost symmetric shapes.
\end{acknowledgments}


\begin{references}
\bibitem{r01}A. Abragam, \textit{The Principles of Nuclear Magnetism} (Oxford University Press, Oxford, 1961).
\bibitem{r02}C.P. Slichter, \textit{Principle of Magnetic Resonance} (Springer-Verlag, Berlin 1992).
\bibitem{r03}D.J. Griffiths, \textit{Introduction to Electrodynamics} (Prentice Hall, Upper Saddle River, New Jersey, 1999).
\bibitem{Ref01}J.D. Jackson, \textit{Classical Electrodynamics}, Second edition (John Wiley \& Sons, New York, 1975).
\bibitem{Ref03}H. Nagaoka, J. Coil Sci. Tokyo \textbf{27}, 18 (1909).
\bibitem{r06}F.W. Grover, \textit{Inductance calculations Working Formulas and Tables} (Dover Publications, Inc. New York 1962).
\bibitem{Ref04}A.B. Pippard, \textit{The Physics of vibration volume 1, containing Part 1, The simple classical vibrator} (Cambridge University Press, Cambridge, 1978).
\bibitem{Ref02}Stanford Research System, SR850 Instruction Manual. 
\end{references}
\end{document}